\documentstyle[emulateapj]{article}
\input epsf.tex
\newcommand{\be}{\begin{equation} }
\newcommand{\ee}{\end{equation}}
\begin{document}
\title{Failed Oort Clouds and Planetary Migration}
\author{Brad M. S. Hansen$^{1}$ }
\affil{Princeton University Observatory,
Princeton, NJ, 08544-1001$^{2}$}
\altaffiltext{1}{Hubble Fellow}
\altaffiltext{2}{email: hansen@astro.princeton.edu}

\begin{abstract}
Planet formation is accompanied by the formation of comet clouds. In systems
where planets migrate on rapid timescales, the diffusive evolution of comet
orbits may stall, resulting in a comet cloud intermediate between a flattened
Kuiper Belt and a spherical Oort cloud. These `failed Oort clouds' may provide
a `smoking gun', indicating that planetary migration has taken place. If some
fraction of the scattered component consists of planetary embryos, it may be
possible to observe transits of such bodies even when the planetary system is
not edge-on to the line of sight.
\end{abstract}

\keywords{planetary systems -- comets:general -- Kuiper Belt -- Oort Cloud
-- planets \& satellites:general -- occultation}

\section{Introduction}

Planet formation is a messy business. The process of accumulation of small
bodies into large bodies results not only in the formation of large planetary-mass
objects but also a significant population of scattered comets and asteroids.
Studies of short and long period comets in our solar system
(Oort 1950; Fernandez 1980; Duncan, Quinn \& Tremaine 1988) point to the existence
of two repositories of cometary material, the spherically distributed Oort cloud
(Oort 1950) and a low inclination population of trans-Neptunian objects, the
Edgeworth/Kuiper Belt (Edgeworth 1949; Kuiper 1951). These cometary reservoirs
convey valuable information regarding the primordial conditions in the solar system
and their manifestations around other stars offer insights into the formation
of extrasolar planetary systems. In particular, the detection of infra-red emission
from dust in such systems (Backman \& Paresce 1993; Trilling \& Brown 1998) is
 thought to supplied
by ongoing evaporation or ablation of comet-like objects (Weissman 1984) .

The properties of recently detected extrasolar planets
(e.g. Marcy, Cochran \& Mayor 1999) suggest that the formation 
and  evolution of planetary systems may be a much more dynamic and violent process
than previously envisaged. The purpose of this letter is to examine the impact of
this new paradigm on the configuration of cometary reservoirs in such systems.
In particular, we will demonstrate the existence of 
a cometary component intermediate between a disk and isotropic component that will
result in systems where significant planetary migration has taken place.

Section~\ref{Scatter} reviews the process of formation of an Oort cloud and examines
the implications of planetary migration for such a scenario. In section~\ref{Dust} we
describe the implications for observations of dust disks around extrasolar planetary
systems.

\section{Planetary Scattering \& Oort Clouds}
\label{Scatter}

Oort (1950) inferred the existence of a spherically distributed cometary reservoir
containing objects of semi-major axis $a \sim 10^4$~Au. Perturbations of cometary
orbits by passing stars or molecular clouds are responsible for scattering comets 
into the inner solar system. The origin of these bodies is explained by ejection
of primordial material from the solar system.
 Proto-comets on planet crossing orbits
are repeatedly scattered in close encounters with the outer planets and their orbits
undergo a diffusive evolution towards large semi-major axis and high eccentricity.
This diffusion continues until they are ejected from the solar system entirely or
until perturbations at apastron due to the Galactic tide perturb the orbit sufficiently
to remove it from a planet-crossing condition (Duncan, Quinn \& Tremaine 1987, Tremaine 1993).
Continued action of the tidal torques eventually result in 
 an isotropically distributed population, responsible for the 
long period comets.

The origins of the short period comets seem incompatible with the properties of
the Oort cloud (Joss 1973; Duncan, Quinn \& Tremaine 1988) and are now thought to
be related to a disk-like distribution of primordial protosolar material 
(Edgeworth 1949; Kuiper 1951; Fernandez 1980) in trans-Neptunian orbits. Recent observations
(Jewitt \& Luu 1993; Stern 1996) indicate the presence of $\sim$~100~km bodies in 
this region and possible true cometary material as well. (Cochran et al 1995)

\subsection{Migration}

 Implicit in the studies of the formation of the Oort cloud 
(e.g. Duncan, Quinn \& Tremaine 1988) is the assumption that
the planets themselves do not move significantly. 
Recent dynamical studies of the early evolution of our own solar system indicate that some
small migration of the outer planets is likely (Fernandez \& Ip 1984; Malhotra 1995; Hahn \& Malhotra 1998).
Recent discoveries of jovian mass planets in very close orbits around other stars
(Mayor \& Queloz 1995; Marcy \& Butler 1996; Marcy, Cochran \& Mayor 1999) suggests that, in some cases,
far more extensive migration occurred. 

The explanations for the source of this migration invoke either some kind of steady
orbital decay (Lin, Bodenheimer \& Richardson 1996; Murray et al 1998) or the scattering
of several planets in an originally unstable orbital configuration (Rasio \& Ford 1996;
Weidenschilling \& Marzari 1996; Lin \& Ida 1998). However, the emphasis on migration
as opposed to in situ formation is still largely theoretical prejudice. As I shall
demonstrate, the Oort clouds/Kuiper belts around these systems could provide a potential
test of the migration hypothesis.

In the standard picture of Oort cloud formation, a comet scattering off the giant planets
returns repeatedly to the inner solar system until either something catastrophic (ejection
or collision with a planet) happens or the orbit receives a sufficiently large external perturbation
that it no longer enters the inner solar system. However, in a system where the scattering
planet(s) are migrating, it is possible that the planet itself moves sufficiently far
between encounters so
that the comet no longer crosses its orbit and is thus no longer subject to strong perturbations.
Thus, the diffusive evolution of comets to large semi-major axes and high eccentricities may
`stall' at much smaller radii than in the standard scenario.
The timescales to eject comets to the Oort cloud is of order $10^6-10^7$ years (e.g. Dones et al 1996), a
timescale similar to that on which migration is claimed to occur (Lin et al 1996; Murray et al 1998).
Thus it is reasonable that planetary migration  will leave many comets
stalled in their diffusion process, on eccentric and inclined orbits of moderate semi-major axis. One possible exception is those scenarios which invoke 
dynamical instability in multi-planet systems, which could occur at much later times
when the asteroids \& comets have been mostly cleared.

Thus, in the case of steady migration, we expect a cometary component intermediate between
a Kuiper belt disk population and an Oort cloud isotropic distribution. The orbital extent and
inclination distribution will depend on the rate of planetary migration as well as other
evolutionary factors, which we will describe below.

\subsection{Cloud Size}

Figure~1 shows the comet cloud that results when one considers the formation of a 51~Pegasi-like system in the planetesimal migration scenario of Murray et al (1998). The planet 
initially begins at 10~Au and migrates to 0.06 Au, leaving behind a cloud of comets or
asteroids in a thickened disk.
 The evolution was followed
using a Monte-Carlo code (Hansen et al, in preparation) utilising the \"{O}pik
approximation (\"{O}pik 1976; Arnold 1965). In this particular instance, approximately
$50 \%$ of the disk planetesimal mass interior to the original
orbit is scattered into a "thick" component. 

While the appearance of such thickened configurations is generic in single planet migration
models, the exact nature of the scattered component will depend on the particular nature
of the system, most particularly, the mass of the migrating planet and the rate of
migration. Smaller planets migrate faster and result in less thickening. The rate of
migration also slows as the planet gets closer in, so that the thickening of the disk
increases inwards. Larger mass planets halt migration at larger radii, so that the 
scattered cloud shows an inner edge. Figure~2 shows the result for a 1~$M_J$ planet that
migrates to 0.17 Au. The asteroid cloud truncates near 1 Au.

The comets scattered into inclined, eccentric orbits will also precess under the influence
of any undisturbed extended disk component exterior to the original planetary orbit, on
timescales $\sim 10^5$ years. The eccentricity and inclination undergo oscillations governed by Kozai's integral $\Theta = (1 - e^2) \cos^2 i$ (Kozai 1962; Holman, Touma \& Tremaine
1997). Thus, the minimum semi-major axis (maximum eccentricity) occurs when the comet
lies in the orbital plane again. A small fraction of the comets at the inner edge of the
cloud may again be brought into planet crossing orbits by this mechanism and ejected. 
However, this is only a small fraction ($\sim 5 \%$) of the scattered component in the
above cases.

The longevity of such a cloud is also affected by the presence of other planets in the
system. Additional planets will also scatter planetesimals and may serve to clear out
orbits left behind by the original migrating planet. Figure~3 illustrates the situation
for two planets starting at 10 and 20 Au. The inner planet migrates down to inside 0.1 Au
as in figure~1 while the second, jupiter mass planet stops at 1.6 Au. The resulting
scattered cloud consists of an inner and outer component. The eccentricities of the
inner cloud components are restricted by the fact that they cannot cross the orbit
of either planet in the final state, otherwise they will be ejected. This is why
there is no intermediate cloud left in our own solar system - the limited movement
of the outer planets meant that, once a body crossed a planetary orbit it was doomed
to continue scattering until it was either ejected or passed into the Oort cloud.

In conclusion, we reiterate that the above calculations represent the situation in the migration
scenario of Murray et al(1998). In this case the migration is intimately coupled to the scattering
of the planetesimals as these provide the sink for the planet's gravitational binding energy.
As such the results are robust to uncertainties inherent in the \"{O}pik approximation
(see Dones et al 1996) as these relate primarily to timescale. A similar result will occur
in gaseous migration scenarios such as described by Lin, Bodenheimer \& Richardson (1996), but the
quantitative results will be more sensitive to the relative rates of migration and cometary orbital
diffusion.

\section{Implications}
\label{Dust}

The search for residual material in known planetary systems is a rapidly advancing field.
Observations of infra-red excesses and transient absorption events in the Beta Pictoris
system have been explained by the existence of cometary bodies (Weissman 1984; Lagrange-Henri et al 1989; Artymowicz \& Clampin 1997; Kalas et al 2000). Extensive studies of the
dust distribution suggests that the properties change at smaller radii (Lecavalier des Etangs, Vidal-Majar \& Ferlet 1996; Li \& Greenberg 1998). While this is a reasonable explanation,
it should be noted that most models assume a disk of constant opening angle and that the
change in the disk properties suggested by the above results can influence the observations
as well. A configuration similar to Figure~2 can 
perhaps explain the inner gap inferred in Beta Pictoris system. The scattered cloud
is also a natural reservoir for bodies suggested to give rise to the transient
absorption events observed in this system (Lagrange-Henri et al 1989). 

Trilling \& Brown (1998) have recently detected infra-red excesses using coronagraphic
techniques around the 55~Cancri planetary system. They
 have used the observed axis ratio of the 55~Cancri disk to infer
the inclination angle of the system, assuming a thin dust disk configuration and thereby
constrain the true planet mass. The presence of a thickened disk as discussed
above will change the inclination and mass
estimates. The observed ratio of minor to major axis $b/a$ is now $\sim \cos (\theta +
\delta \theta)$, where $\theta$ is the inclination angle and $\delta \theta$ the disk                            thickness. Thus, an appreciably fattened disk will cause an overestimate of the inclination
angle and an overestimate of the planetary mass. A significant fraction of the $\sim 27^{\circ}$ inclination meas
ured by Trilling \& Brown could be the result of disk thickening
(see Figure~1).    

The larger members of these close, approximately spherical clouds may be detectable
by transit observations. Our knowledge of the Kuiper Belt is due to the largest rocky
bodies rather than the presumed cometary component. Similarly, the first detected objects in
these failed Oort clouds may not be cometary but rather large asteroids or planetary embryos.
 The formation of Jovian cores may also leave a substantial population of
planetary embryos.
 If so, one would expect to see transits of
such bodies in many systems with close-in Jupiters, since they are not confined to
the planetary orbital plane. To indicate this, consider the simulation shown in Figure~1.
If we consider the scattered objects to be protoplanetary embryos $\sim 0.2$ earth masses
and radii $R \sim 4000$~km, then the instantaneous geometrical filling factor for
inclinations larger than 5 degrees out of the planetary orbital plane $\tau_1 = \sum_i \pi
(R_i/a_i)^2 \sim 6 \times 10^{-5}$. However, for continuous monitoring over an
orbital period, the optical depth to see a transit (with fractional depth $\sim 10^{-4}$
in this case) is $\tau_2 = \sum_i \sin \theta R_i/a_i \sim 0.09$.
 Although these numbers
are quite uncertain because of the unknown properties of the scattered bodies, they do
indicate the desirability of re-examining {\em all} the known systems for transits
as the technology improves, including those in which edge-on orbits for the known planets
have been already ruled out.

In conclusion, the presence of a cometary or asteroidal component intermediate between
the Kuiper Belt and the Oort cloud is a potential indicator of significant planetary
migration. Such a component may also provide an important ingredient in the studies of
young protoplanetary disks.

Support for this work was provided by NASA through Hubble Fellowship grant \#HF-01120.01-99A,
from the Space Telescope Science Institute, which is operated by the Association of Universities
for Research in Astronomy, Inc., under NASA contract NAS5-26555.



\figcaption[ai2.ps]{The distribution of inclinations for a planetesimal disk 
through which a planet has migrated to small radii. The inset also shows the eccentricity
distribution. The labels `initial' and `final' indicate the initial and final orbital
semi-major axis of the planet.
}

\figcaption[ai4.ps]{As before, we show the inclination distribution of the remnant
planetesimal disk. In this case the planetary migration stalled at 1.6~Au, leading to
an inner gap in the planetesimal disk.}

\figcaption[ai3.ps]{The inclination distribution for a system of two planets. This
results in both an inner and outer thick disk, with a gap $\sim$ several Au, in
between.}

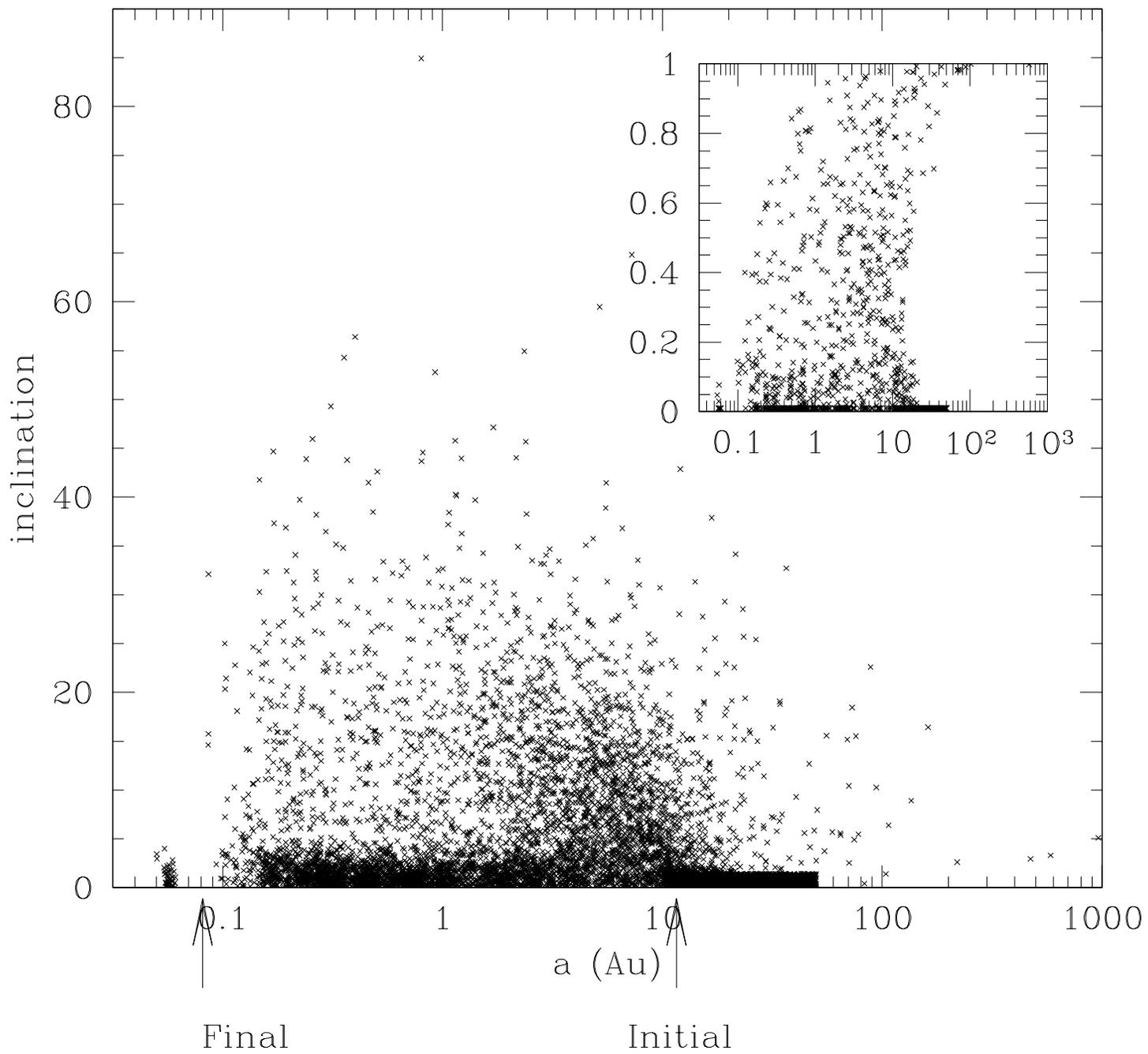
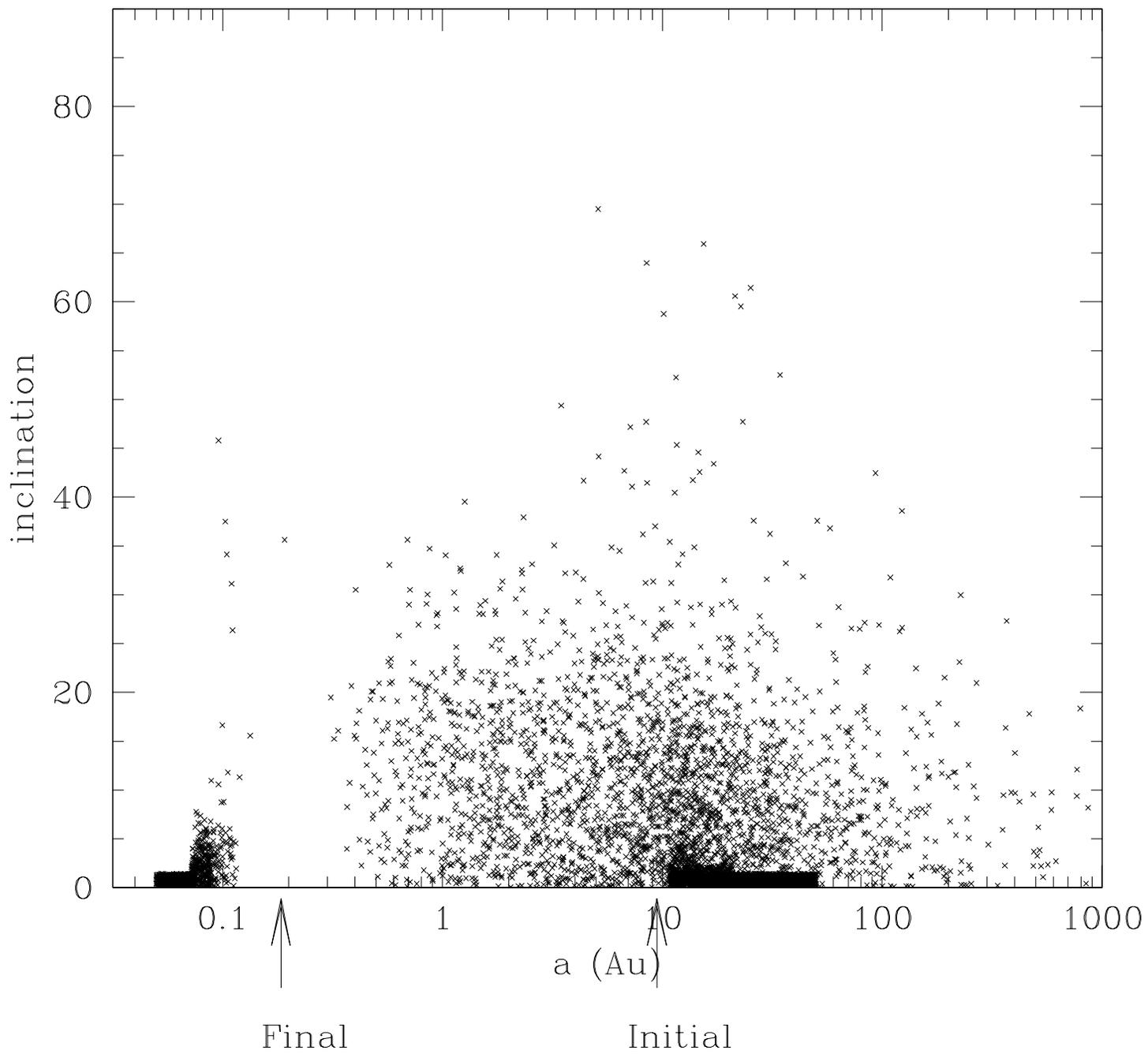
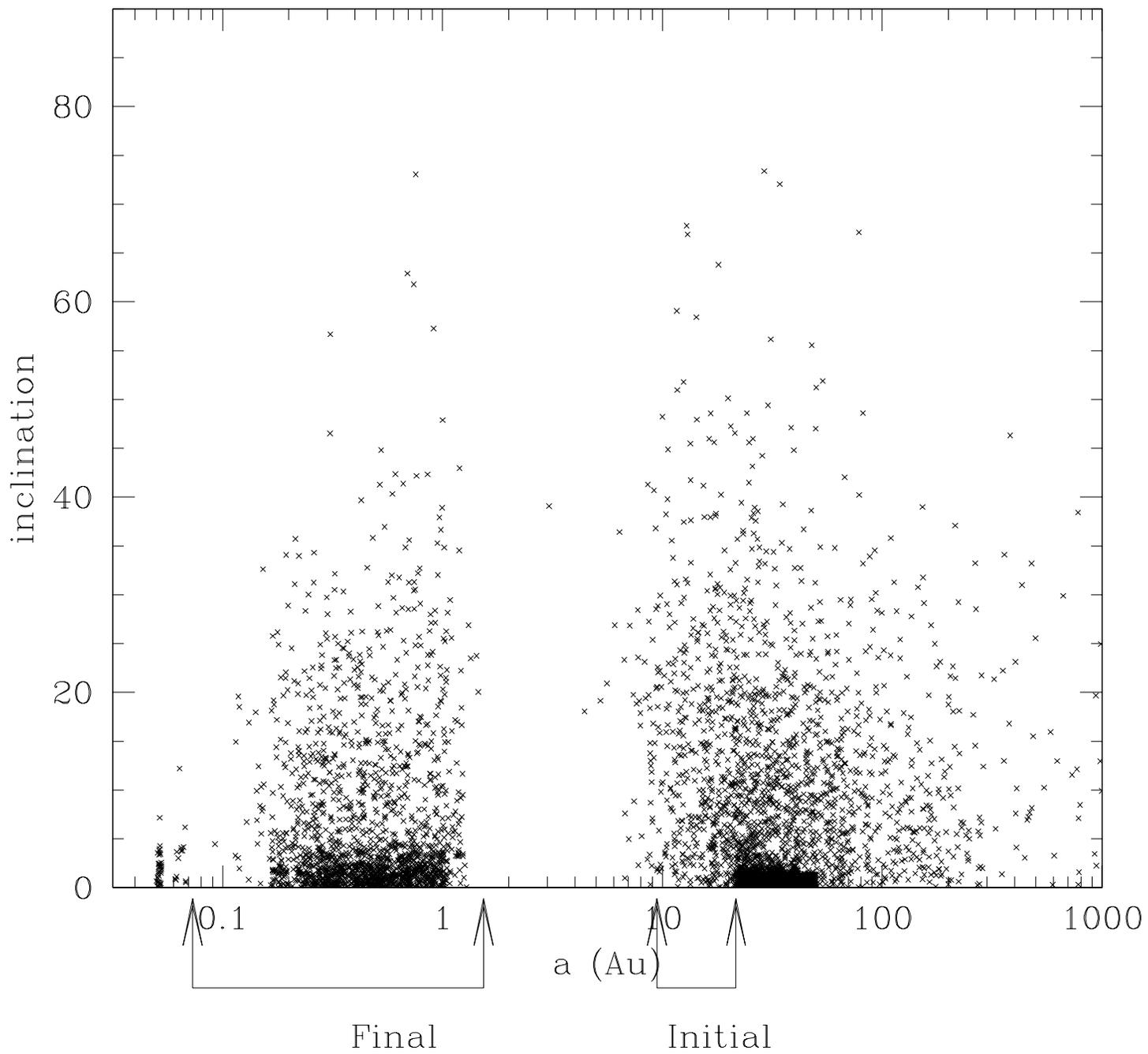
\end{document}